\documentclass[epj]{svjour}
\usepackage{latexsym}
\usepackage{graphicx}
\usepackage{amsmath,amssymb}
\begin{document}


\title{Investigation of Quantum Phase Transitions using Multi-target DMRG Methods}
\author{C. Degli Esposti Boschi\inst{1} \and F. Ortolani\inst{2}}

\institute{Unit{\`a} di ricerca INFM di Bologna, \email{desposti@bo.infm.it},\\
viale Berti-Pichat, 6/2, 40127, Bologna, Italia \\
\and
Dipartimento di Fisica, Universit{\`a} di Bologna, INFM and INFN, \email{ortolani@bo.infn.it},\\
via Irnerio, 46, 40126, Bologna, Italia}


\abstract{
In this paper we examine how the predictions of conformal invariance
can be widely exploited to overcome the difficulties of the density-matrix
renormalization group near quantum critical points. The main idea is to match
the set of low-lying energy levels of the lattice Hamiltonian, as a function
of the system's size, with the spectrum expected for a given conformal field theory in two dimensions. 
As in previous studies this procedure requires an accurate targeting of various excited states.
Here we discuss how this can be achieved within the DMRG algorithm
by means of the recently proposed Thick-restart Lanczos method. As a nontrivial 
benchmark we use an anisotropic spin-1 Hamiltonian with special attention
to the transitions from the Haldane phase. Nonetheless,
we think that this procedure could be generally valid
in the study of quantum critical phenomena.}


\authorrunning{C. Degli Esposti Boschi and F. Ortolani}
\titlerunning{Multi-target DMRG Methods for Quantum Phase Transitions}

\maketitle

\section{Outline and General Facts}
\label{S_Outline}
\par The density-matrix renormalization group (DMRG) was invented by S. R. White in the early 90's
and nowadays is recognized as one of the most accurate and efficient numerical techniques
in the study of correlated quantum systems \cite{PWKH}. A number of authors is still
contributing to its development for new applications and one of the major points is
the dimensionality of the system. While for 1D lattices of spins or electrons there exist
several firm results, a generally accepted DMRG scheme in 2D is still lacking despite
numerous proposals. However, even for 1D systems, it is known that the DMRG encounters
some difficulties in reproducing the physics of quantum critical points.
This is because one wants, at the same time, to explore larger and larger system sizes
and to maintain a good numerical precision in order to locate the points where the excitation
gap closes and to compute the associated critical exponents. If the quantum chain has
$q$ states per site, the dimension of the full Hilbert space for $L$ sites grows exponentially
as $q^L$. The main approximation of the DMRG is the truncation to the space
spanned by the $M$ eigenvectors of the block density matrix, $\rho_{\rm b}$, corresponding
to the $M$ largest eigenvalues. The problem is that, close to criticality,
the system experiences fluctuations on very large scales and (as will be discussed
in sec. \ref{S_DepM}) the spectrum of $\rho_{\rm b}$ decays more slowly than in noncritical
cases. Hence if one considers, as a first indicator of the accuracy of the method,
the sum of the discarded weights $W_M = \sum_{j > M} w_j$, then this different decay
implies that the minimal $M$ to reach a prescribed threshold is generally larger when the system
is close to a critical point.  
To be more specific, Legeza and co-workers \cite{LF,LRH} found that the error in the
ground state (GS) energy due to a finite $M$, at a fixed $L$,
is simply:
\begin{equation}
\delta E_L^{\rm GS}(M) = E_L^{\rm GS}(M) - E_L^{\rm GS}(\infty) \propto W_M \; .
\label{LfeGSe}
\end{equation}
Now, if we admit that the precision for the first excited state follows a similar
trend, then there will be a regime, close to the critical point and/or for large $L$
(see sec. \ref{S_IdCFT}), where $W_M$ becomes larger than the excitation gap
itself. In addition, figs. 3 and 5 of ref. \cite{LF} indicate two other important
features (at least for the Ising model in a transverse field).
First, at criticality the error in the energy at fixed $M$ increases appreciably with
the chain's length, and again we see that fixing $M$ once for all, while taking larger and larger
values of $L$, is not a completely safe procedure. 
The second, and more important point is that the proportionality in eq. (\ref{LfeGSe}) holds
only after a few {\it finite-system} iterations have been performed. 
Using Legeza's terms, these are necessary to cancel the so-called environment error
that dominates at sufficiently small $L$. The crossover from the environment-dominated
regime to the truncation-dominated one is marked by a characteristic size, $L^*(M)$,
which increases with increasing $M$. In sec. \ref{S_ER} we will argue that a similar
characteristic length emerges in critical spin-1 chains too. Moreover, in our opinion,
the results of Andersson, Boman \& {\"O}stlund \cite{ABO} regarding a critical chain of free spinless
fermions, or equivalently a spin-1/2 XX model, point somehow in the same direction.
Even if the system is known to be rigorously critical, the effect of a finite number of DMRG
states is to introduce a {\it fake correlation length}, that grows as $M^{1.3}$.
The interplay between an analogous DMRG length and the true correlation length near the
critical temperature of classical 2D systems is also discussed in \cite{NOK}.  

\par We believe that the considerations above indicate that the critical behavior of (1D) quantum systems
cannot be ``simply'' approached by using the computational resources to reach larger
and larger values of $L$ by means of the infinite-system DMRG, as it is sometimes
done (see, for instance, \cite{CCC} and refs. therein). Rather, we prefer to exploit as much as
possible the consequences of {\it conformal invariance}. In many cases, indeed,
we expect that
the low-energy physics of a 1D quantum system near its critical points is described by
a suitable conformal field theory (CFT) in (1+1) dimensions \cite{Ko,ZF,PKO,OHA}.
Then, if the system under study has to be critical, we imagine that there will
be a scale-invariant effective continuum model that captures its universal features.
In 2D the key point is that all these universality classes are
encompassed in the framework of CFT. 
In the last twenty years the numerous exact results in this area have been organized into an elegant
framework, which is however too vast for our purposes. Therefore, in sec. \ref{S_IdCFT}
we sketch only the points of contact with our analysis and refer to \cite{DFMS} for
a comprehensive guide and to \cite{Ra} for a shorter review on finite-size effects.

\par However, for a correct interpretation of the DMRG data, and a comparison with a suitable
CFT, an accurate calculation of the first excited states at varying $L$ is needed.
The targeting of more than one state was proposed by White himself \cite{Wh93}
and used by different authors in subsequent papers. A recent comment on this problem can be
found in \cite{Bu}. One of the aims of our work is to test, in connection with the DMRG,
the recently proposed Thick-restart Lanczos method \cite{WS} to handle a sufficient
number of excited states. 

\section{Identification of the CFT through the Finite-Size Spectrum}
\label{S_IdCFT}
\par The requirement of scale invariance in 2D is sufficiently strong to
allow (under general assumptions) a classification of the states of a quantum field theory in
terms of the irreducible representations of the Virasoro algebra generated by a set
of operators satisfying:
\begin{equation}
[L_m,L_n]=(m-n)L_{m+n}+\frac{c}{12}\delta_{m+n,0}(m^3-m) \; , \;\; m,n \in {\mathbb Z}
\label{Vir_alg}
\end{equation}
(As is customary in CFT, the statements and equations regarding the holomorphic part should
always be suitably repeated for the antiholomorphic part, denoted by overbars).
For unitary theories the central charge of the algebra is either $c \geq 1$, or can be chosen as one of the 
values $c=1-6/p(p+1)$ ($p=2,3,\dots$), each one corresponding to a so-called
minimal model. In the latter case, once a $c < 1$ is given,
one can decompose the Hilbert space into a {\it finite} number of irreducible representations
of (\ref{Vir_alg}) labeled by the eigenvalues of the generator $L_0$:
\begin{equation}
L_0 \vert \Delta \rangle = \Delta \vert \Delta \rangle \; , \;\; \Delta=\frac{[(p+1)r-ps]^2-1}{4p(p+1)} \; ,
\label{ev_L0}
\end{equation}
with
$1 \leq s \leq r \leq p-1 \; \in {\mathbb Z}$.
More specifically, the primary states $\vert \Delta \rangle$ are annihilated by $L_m$ with positive $m$ and each one of them 
generates a conformal family that
contains all the states obtained through the application of the negative-integer generators
$\vert \Delta \rangle_m^k = (L_{-m})^k \vert \Delta \rangle$ (the secondary or descendant states). 
It can be seen, as a consequence of the Virasoro algebra, that these are in turn eigenvectors
of $L_0$:
\begin{equation}
L_0 \vert \Delta \rangle_m^k = (\Delta+mk) \vert \Delta \rangle_m^k \; . 
\label{ev_L0_mk}
\end{equation} 
In principle, all the correlation functions 
of the theory can be reduced to correlators of primary operators. 
Moreover, the operators $L_0$ and $\bar{L}_0$ play a special physical
role as can be seen by considering a conformal transformation that maps the plane onto
an infinite cylinder, whose circumference of length $L$ represents the space axis with periodic
boundary conditions (PBC).
Then the energy and momentum operators are simply expressed as:
\begin{equation}
H_{\rm CFT}=\frac{2 \pi v}{L}\left[L_0+\bar{L}_0-\frac{c}{12}\right] \; , \;\;
Q=\frac{2 \pi}{L}\left[ L_0-\bar{L}_0 \right] \; .
\label{HCFT_Q}
\end{equation}
Actually, the pre-factor $v$ has been inserted ``by hand'' in view of the identification
of the (critical part of the) spectrum of the original quantum Hamiltonian 
with the values predicted by the CFT construction. More specifically,
the ground state is simply identified with $\vert {\rm vac} \rangle$, for which both $L_0$ and
$\bar{L}_0$ have null eigenvalues. Thus, the finite-size corrections to the vacuum energy
are:
\begin{equation}
E_L^{\rm vac}= - \frac{\pi c v}{6 L} \; ,
\label{Evac_and_c}
\end{equation}
whereas the excited states follow from the construction of eqs. (\ref{ev_L0}) and (\ref{ev_L0_mk}):
\begin{equation}
E_L(\Delta,m,k;\bar{\Delta},\bar{m},\bar{k})-E_L^{\rm GS} = \frac{2 \pi v}{L}[\Delta+\bar{\Delta}+mk+\bar{m}\bar{k}] \; .
\label{EES}
\end{equation}
It is important to recall that the sum of the eigenvalues of $L_0$ and $\bar{L}_0$, that is
the last term in square brackets $d_O \equiv (\Delta+\bar{\Delta}+mk+\bar{m}\bar{k})$,
is called the {\it scaling dimension} because it enters the
algebraic decay of the correlation function of the corresponding operator $O$:
\begin{equation}
\langle O(0,0) O(z,\bar{z}) \rangle \sim z^{-2(\Delta+mk)} \bar{z}^{-2(\bar{\Delta}+\bar{m}\bar{k})} 
\vert_{z=\bar{z}=r} = r^{-2d_O} \; .
\label{O_and_itsd}
\end{equation}
In the language of renormalization group (RG) theory, the operators are seen
to be relevant when $d_O < 2$, irrelevant when $d_O > 2$ and marginal when the scaling dimension is exactly 2
(in 2D). The exponents, $y_i$, of the scaling fields near a fixed point of the RG flow
are simply given by $y_i=2-d_i$. 

\par The edge case $c=1$ opens the way to unitary CFT with an infinite number of primary operators.
It is related to certain topological constructions of a free bosonic field \cite{Gi}, and the operators
(or the associated states) can be further classified in terms of an $\hat{U}(1)$ Kac-Moody algebra.
Here we won't enter into more details since
ref. \cite{DEBEOR} is specifically devoted to the study of such $c=1$ CFT starting from the spin-1 
Hamiltonian (\ref{HlD}), using just the method explained in the present paper.
Finally, when $c>1$ we have a {\it continuum} of allowed values and
the complexity of the problem in its generality simply forbids us to go on with the discussion
of classified cases. We refer to the literature on CFT for general discussions (see, for example,
\cite{DFMS,Gi}).
We will content ourselves with the observation that, at least from an operative point of view, 
in many cases eqs. (\ref{HCFT_Q})-(\ref{O_and_itsd}) should remain valid once the labels $\Delta, 
\bar{\Delta}$ and the secondary indices are
properly interpreted in terms of wider symmetries that have to be investigated case by case.

\par The first problem is, of course, that one does
not know, {\it a priori}, which is the CFT that has to be invoked.
A great step towards the answer to this question is made if one knows the
central charge of the theory, $c$, that characterizes the underlying
Virasoro algebra. 
Apart from eq. (\ref{Vir_alg}) and the operator product expansion of the stress-energy
tensor, the central charge appears explicitly in the size-dependence of the GS energy density:
\begin{equation}
\frac{E_L^{\rm GS}}{L}=e_\infty - \frac{\pi c v}{6 L^2} \; ,
\label{EGS_and_c}
\end{equation}
which is nothing but eq. (\ref{Evac_and_c}) plus an infinite-size term, which is absent in
the formal CFT but has a definite value for a given quantum Hamiltonian.
If the quantum chain has open boundary conditions, which are generally better for DMRG convergence,
eq. (\ref{EGS_and_c}) should be modified in two ways \cite{BCN,TM}. The denominator becomes
$24 L^2$ and a boundary term $B/L$ should be added, with a non-universal
pre-factor $B$. This term introduces a slower convergence of the GS to the thermodynamic limit (TL) 
and, at least in this sense, PBC may be more useful for finite-size scaling
(FSS) \cite{BMcKH}.

\par In quantum field theory the second term of (\ref{EGS_and_c}) accounts for the Casimir effect \cite{BCN},
while in statistical systems it gives the correction to the free energy at small
temperatures ($L$ playing the role of $1/T$ \cite{Af}). 
In our case, eq. (\ref{EGS_and_c}) is the starting point to discover the CFT
that is appropriate for the problem under consideration.
Indeed, having good estimates of $E_L^{\rm GS}$ at various $L$, we can first best-fit
$e_\infty$ and $c v$. Then, if we have reasons to believe that one of the states
has scaling dimension 1 (eqs. (\ref{EES}) and (\ref{O_and_itsd})), then we can calculate:
\begin{equation}
v=\frac{\Delta E_L(d=1)}{2 \pi/L} \; .
\label{v_as_dEdk}
\end{equation}
If we now interpret $\Delta k=2 \pi/L$ as the quantum of momentum
for a chain of length $L$, then eq. (\ref{v_as_dEdk}) looks like a discrete
derivative of the energy vs momentum relation, that is, a group velocity.
This somehow explains the term ``spin velocity'',
which is widely used in the literature of quantum spin chains independently
of the real nature of the excitations. Again, the exact form of eq. (\ref{v_as_dEdk}) may vary
depending on the boundary conditions. In spin chains with twisted boundary
conditions the general relation in eq. (\ref{EES}) remains valid
and it's the scaling dimensions of the operators that turn out to depend
on the twisting angle \cite{Ki}. With open boundary conditions,
according to ref. \cite{BCN} the essential
modification is the replacement of
$L$ by $2L$. This justifies the coefficient of $L^2$ in eq. (\ref{EGS_and_c})
and eq. (58) of \cite{TM}, that looks like eq. (\ref{v_as_dEdk}) with
a quantum of momentum $\Delta k=\pi/L$. 

\par Summing up, with a combined usage of eqs. (\ref{EGS_and_c}) and (\ref{v_as_dEdk})
we can calculate $c$ from the numerical data of the first excited levels
for different chain's lengths. Then the identification may fall into two cases.
For $c < 1$, unitarity demands that the values of the central charge are
quantized, according to the list of minimal models. In this case, a small
discrepancy from one of these values is likely to be related to numerical
uncertainties. The matter becomes more complicated when one finds a numerical
$c$ larger than one, in which case unitarity by itself is not enough
to provide a quantization condition. The symmetries of the lattice Hamiltonian
in this case are of great help because we expect them to be present
also in the corresponding continuum model. Then one may focus on the known
2D field theories whose actions are invariant under both this symmetry
group and under conformal transformations. Beside that, when one studies a
novel and computationally demanding system it may be very useful to
get a  first insight of the CFT by performing a finite-size
analysis of DMRG data obtained with the infinite-system algorithm and open boundary conditions, as in \cite{TM}.
Nonetheless, we discourage a blind usage
of the infinite-system DMRG to conclude (as in ref. \cite{CCC}) that the
numerical discrepancies from the expected values are to be ascribed
to nontrivial effects beyond the CFT framework. If in doubt, the
final answer should always come from a careful refinement using {\it finite-system}
numerical data, in a range of $L$ where the scaling behavior is visible
but the accuracy is not corrupted significantly by the DMRG truncation error.

\par In any case, what we are actually performing is a self-consistent guess
of the underlying CFT. In principle, this allows an analytical calculation
of the scaling dimensions, $d_\ell$, of all the operators associated with the excited states.
The corresponding energy gaps, here formally indexed by $\ell$, scale according to eq. (\ref{EES}):
\begin{equation}
\Delta E_L^\ell = \frac{2 \pi}{L} v d_\ell \; .
\label{EaES}
\end{equation}
Matching the numerical spectrum of a certain number of levels 
with the structure encoded in the scaling dimensions
of eq. (\ref{EaES}) represents a particularly stringent test of the hypothesis made.

\section{Model and Details of the Implementation: 
\\Multi-target Method}
\label{S_MtM}
\par Following our previous paper \cite{DEBEOR} we have considered the following 
spin-1 Hamiltonian:
\begin{equation}
H=\sum_{j} \left[ \frac{1}{2}\left( S_j^+ S_{j+1}^- + S_j^- S_{j+1}^+ \right)+
\lambda S_j^z S_{j+1}^z + D(S_j^z)^2 \right]
\label{HlD}
\end{equation}
for a chain of $L$ spins which includes both an Ising-like and a single-ion anisotropy term, 
with coefficients $\lambda$ and $D$ respectively. 
Formally we have set $\hbar =1$ and the overall coupling constant $J=1$
so that every quantity turns out to be dimensionless and we have imposed 
PBC: ${\bf S}_{j+L}={\bf S}_j \;\; \forall j =1,\dots L$. 
See \cite{DEBEOR,BJK,GS,Sc,KT,CHS} for a discussion of the various phases in the $\lambda$-$D$ 
diagram.

\par The algorithm that we have implemented for DMRG calculations follows rather closely
the lines indicated by White in his seminal papers \cite{Wh93,Wh92}. 
As regards our specific application, we should outline the following points.

\par The superblock geometry was chosen to be $[{\sf B}^r \bullet
\vert {\sf B}^{r'}_{\rm ref} \bullet]$ with PBC, 
where ${\sf B}^{r'}_{\rm ref}$ is the 
(left $\leftrightarrow$ right) reflection of block ${\sf B}^{r'}$ with $r'$ sites.
The rationale for adopting this configuration is that, being effectively on a ring, the two blocks
are always separated by a single site, for which the operators are small matrices that are treated
exactly (no truncation) \cite{Wh93}. In this way we expect a better precision in the correlation functions calculated
by fixing one of the two points on these sites and moving the other one along the block. 
In a recent application of the DMRG to quantum chemistry calculations \cite{LRH}
it has been pointed out that the configuration of the superblock may be one of the
major points of optimization of the method for future applications.
As regards the choice of the boundary conditions, we are aware of the fact that
with open conditions a smaller $M$ is generally required and that in certain
cases (i.e. Gaussian transitions) the introduction of twisted boundary conditions is a clever trick to identify
the critical point using numerical data on relatively small systems 
\cite{Ki,CHS,KN,CHS_JPSJ}. Nevertheless,
in our set of calculations we have adopted PBC to get rid of the
edge effects that in some cases mask almost all the information contained
in very short-ranged (string) correlation functions. Moreover, we are primarily
interested in the transitions from the Haldane phase, for which it is known
that the finite-size GS acquires a fourfold degeneracy due to the two free effective spins
at the ends of the chain \cite{KT}. With PBC instead, one has a unique finite-size GS
so that the convergence of the Lanczos procedure is better and the analysis
of the numerical gaps is simpler.

\par We used the finite-system algorithm with three iterations. This prescription should
ensure the virtual elimination of the so-called environment error \cite{LF}, which is expected to dominate in the
very first iterations for $L < L^*(M)$ (see below). In fact, it is only after a suitable
number of such sweeps that we may expect that the error in the energies has been
minimized (for a given $L$ and $M$) and eq. (\ref{LfeGSe}) holds true.
Normally the correlations are computed at the end of the third iteration,
once the best approximation of the GS is available.

\par Dealing with quantum spin chains,
we always exploit the conservation of the $z$-component of the total spin,
$M^z$. Typically we are interested in nonmagnetic GS's, that is, with $M^z=0$.
In every studied case the correlations have been calculated
targeting only the lowest-energy state within this sector.
However, in order to analyze the energy spectrum,
we had to target the lowest-energy state(s) in the other sectors
$\vert M^z \vert=1,2,\dots$ and/or target also a few excited states within
the $M^z=0$ sector, depending on the phase under study. 
The standard Lanczos algorithm gives with enough precision the ground state 
of the system, but it is not so accurate for the excited states. Moreover, 
as a consequence of a general theorem on tridiagonal symmetric matrices, we 
can't have degenerate states from this method. So the algorithm must be modified 
to allow the building of the reduced density matrix over the 
block as a mixture of the matrices corresponding to each target state. 
While for the latter point we are not aware of any specific ``recipe'' 
other than that of equal weights, the implementation of the multi-target 
diagonalization routine within our DMRG code is based on the  
Thick-restart Lanczos method recently introduced by Wu and Simon \cite{WS}. 
In principle, the strategy used here does not rely on a particular
algorithm to extract a group of eigenvalues of the superblock Hamiltonian.
The methods commonly used so far are the Davidson-Liu and the block Lanczos.
Our choice of the Thick-restart Lanczos was based on the intention of testing and exploiting the stability
of the method, as claimed by the authors \cite{WS}. Moreover, its implementation is
a simple extension of the standard Lanczos procedure, as described in appendix A.

\par Once $M^z$ is fixed, in a given run we wish to follow simultaneously the
first levels $\vert M^z ; {\tt b} \rangle$ with {\tt b}=0,1,2,$\dots$,{\tt t} 
(the GS being identified by $(M^z=0;${\tt b}$=0)$). 
Then, as in the conventional Lanczos scheme,
we have to iterate until the norms of the residual vectors
and/or the differences of the energies in consecutive steps
are smaller than prescribed tolerances ($10^{-9}-10^{-12}$ in our calculations).
The delicate point to keep under control
is that, once the lowest state $\vert M^z;0 \rangle$ is found, if we keep iterating searching for
higher levels the orthogonality of the basis may be lost, just because
the eigenvectors corresponding to these levels tend to overlap again with the vector
$\vert M^z;0 \rangle$. As a result, the procedure is computationally more demanding because one has to re-orthogonalize the basis from time to time.
We have seen that this part takes a 10-20\% of the total time spent
in each call to the Lanczos routine. We have also observed that if this re-orthogonalization
is not performed, one of the undesired effects is that the excited doublets (typically due
to momentum degeneracy) are not computed correctly. More specifically, it seems that while 
the two energy values are nearly the same in the asymmetric stages of the sweeps, 
when the superblock geometry
becomes symmetric ($r=r'$ in the notations of the preceding point) the double degeneracy
is suddenly lost in a spurious way. This is in line with the results of ref. \cite{LRH},
where the error in the energy is kept under control by means of a dynamically-adjusted $M$
and the configurations that require a larger number of DMRG states are just those near
the symmetric one.

\section{Dependence on the Number of DMRG States}
\label{S_DepM}
\par As discussed in the introduction of sec. \ref{S_Outline}
the choice of the number of optimized states with respect
to the chain length $L$ is the crucial point to address in any DMRG calculation.
Being conscious that the energy accuracy is {\it not} an exhaustive indicator,
from eq. (\ref{LfeGSe}) one should try, at least, to keep the discarded weight $W_M$
as small as possible. This quantity, in turn, is related to the decay of the density matrix
eigenvalues $\{w_j\}$ as a function of the index $j$:
\begin{equation}
W_M = \sum_{j > M} w_j = \sum_{j} D_j z_j \; ,
\label{disc_W}
\end{equation}
where the last equality is a simple rewriting in terms of the degeneracies
$D_j$ and of the distinct eigenvalues $z_j$.
The issue appeared to be important from the very beginning 
and is the core to the success of the DMRG. In \cite{Wh93,WM} it was argued that the convergence
of the GS energy of a gapped or a spatially finite system 
is roughly exponential in $M$:
\begin{equation}
\delta E_L \propto {\rm e}^{-M/M^*(L)} \; ,
\label{defMastL}
\end{equation} 
with a superimposed step-like behavior, probably related to the successive inclusion of more and more
complete spin sectors. It is also generally believed that this exponential
decay becomes slower (possibly algebraic) when a critical point is approached.
At this stage it is interesting to recall that for integrable systems the
spectrum of $\rho_{\rm b}$ can be determined exactly.
In fact, it is known \cite{NOK,PKO} that for a quantum chain the infinite-size
block density matrix is given by $\rho_{\rm b}=\chi^4$, 
where $\chi$ is the corner transfer matrix of the associated 2D classical statistical system.
This statement has a wide generality, with the exception of critical cases
where the boundary effects may have some role and are expected to affect the tails
of the distribution \cite{CP}. For integrable systems a further step can be made:
$\chi$ is expressed as the exponential of a pseudo-Hamiltonian, $K$, that involves
the same local operators of the Hamiltonian (e.g. $S_j^\alpha S_{j+1}^\alpha$)
but with coefficients depending on the site index $j$. Moreover, the spectrum
of $K$ can be determined exactly and, typically, the eigenvalues turn out to be equally
spaced.
Therefore the distinct eigenvalues of $\rho_{\rm b}$ decay as:
\begin{equation}
z_j \propto Z^j \; , \;\; Z = {\rm e}^{-\epsilon} \; ,
\label{wj_vs_epsilon}
\end{equation}
$\epsilon/4$ being the level spacing of $-K$. These predictions have been explicitly
verified for the Ising model in a transverse
field \cite{PKO} and for the XXZ spin-1/2 Heisenberg chain \cite{PKO,OHA}.

\par Now we shall try to elucidate this topic in the case of critical spin-1
systems using the Hamiltonian of eq. (\ref{HlD}).
At $(\lambda=1,D=0)$ one has an isotropic anti-ferromagnetic (AF) Heisenberg
chain of integer spin, whose theoretical interest comes from what is now known 
as Haldane's conjecture \cite{Ha},
that predicts a genuine quantum behavior with a finite energy gap in the
excitation spectrum. This has to be contrasted with what happens in
half-odd-integer cases that are gapless (i.e. critical) in the thermodynamic limit. 
Another important theoretical contribution
is the mapping between spin chains and restricted solid-on-solid models
proposed by den Nijs and Rommelse \cite{dNR}. Qualitatively, the Haldane phase
is interpreted as a spin-liquid, in the sense that the effective particles
- the spin state $\vert 0 \rangle$ represent an empty site and $\vert \pm \rangle$ represent a
particle with spin up or down -  
are positionally disordered but carry anti-ferromagnetic order.
In the quantum states that contribute to the GS,
this order is hidden by arbitrarily long strings of $\vert 0 \rangle$'s 
but can be measured by the string order correlators \cite{KT,dNR}:
\begin{equation}
O_{\rm S}^\alpha(j,k) \equiv - \left\langle S_j^\alpha 
\exp{ \left( {\rm i} \pi
\sum_{n=j+1}^{k-1} S_n^\alpha \right) } S_k^\alpha \right\rangle \; ,
\label{SOPalpha_cf}
\end{equation}
(the expectation value being taken on the GS) and their asymptotic
order parameters, $O_{\rm S}^\alpha \equiv \lim_{\vert j - k \vert \to \infty}
O_{\rm S}^\alpha(j,k)$.

\par Beside that, nowadays there exist several experimental realizations
of these anisotropic spin-1 chains: to the author's knowledge, RbNiCl$_3$
\cite{BJK} and Y$_2$BaNiO$_5$ \cite{Xuetal,DR} are examples of pure Haldane systems, CsNiFe$_3$ is
a ferromagnet ($\lambda=-1$) with appreciable single-ion anisotropy
($D \simeq 0.4$) and CoCl$_2 \; 2 \;$H$_2$O behaves as an Ising ferromagnet
($ \vert \lambda \vert \gg 1$)
with high easy-axis anisotropy ($D \simeq -5$) \cite{GS,SZ}.
The so-called NENP \cite{RZRV,ZDBR} and NENC \cite{Oretal}
represent, respectively, small-$D$ ($\simeq 0.2$) and large-$D$ ($\simeq 7.5$) antiferromagnets
with easy-plane anisotropy.

\par In order to study the convergence of truncation errors close to criticality
we have selected a pair of representative points in the phase
diagram.
In the origin $(\lambda=0,D=0)$, where a Berezinskii-Kosterlitz-Thouless (BKT)
transition is expected \cite{JP}, we have the spin-1 analogue of the XX spin-1/2 model
used in ref. \cite{ABO} to examine the effect of a finite $M$ on a quantum critical
system. 
The second point that we will examine in this section is $(\lambda=1,D=0.95)$ because
from refs. \cite{DEBEOR,CHS_JPSJ}
we know that it lies very close to the line of transition from
the Haldane phase to the so-called large-$D$ phase. 
In the following numerical analysis these two points will be denoted by XX and H-D, respectively.
The results can be thought as the critical counterpart of the ones presented
in \cite{OHA} for an isotropic spin-1 Heisenberg chain.

\par
\begin{figure}
\begin{center}
\resizebox{0.4\textwidth}{!}{\includegraphics[keepaspectratio,clip]{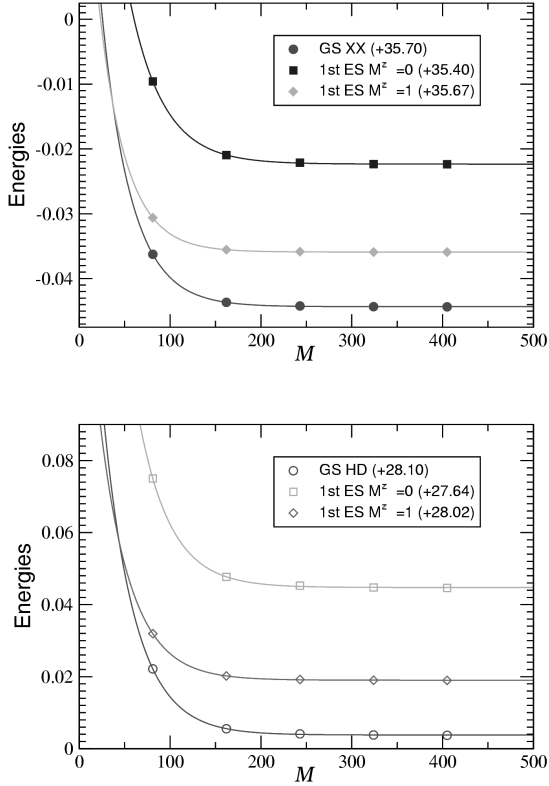}}
\caption{\small Energies of the GS and of the first excited states within $M^z=0,1$ at the points
XX and H-D (see text). The symbols represent
DMRG values obtained with $L=32$ and an increasing number of DMRG states while the continuous lines
are exponential fits. For clarity reasons, the offsets reported in the legend have been added.} 
\label{EnergL32_vs_m}
\end{center}
\end{figure}

\begin{figure}
\begin{center}
\resizebox{0.45\textwidth}{!}{\includegraphics[keepaspectratio,clip]{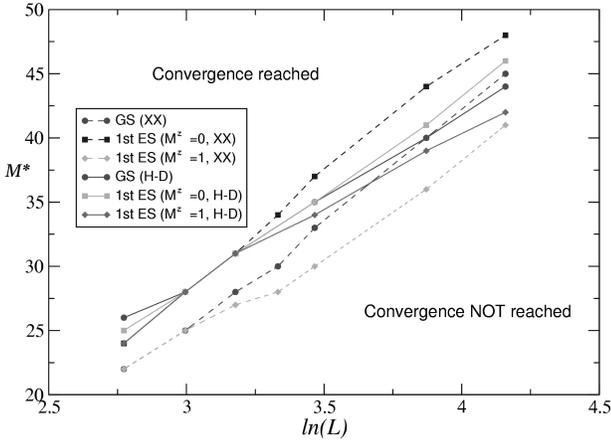}}
\caption{\small Exponential factor $M^*(L)$ (eq. (\ref{defMastL})) for the points indicated 
in the legend (see text). The lines separate
two regions where the convergence of the low-lying levels with $M$ is essentially
reached or not.} 
\label{DMRGcl}
\end{center}
\end{figure}

 For the H-D point we have computed the
first excited state in the sectors with $M^z=0,1$ for $L=16,20,24,32,48$ and 64,
using $M=81,162,243,324$ and 405 DMRG states for each case. For the XX point,
we have monitored the same states
for the same values of $M$ using also $L=28$. 
The exponential decay of eq. (\ref{defMastL}) seems to be 
adequate in all the cases, as shown in fig. \ref{EnergL32_vs_m}.
For every fixed $L$ we best-fit the energy-vs-$M$ data and read off
the characteristic values $M^*(L)$, represented in fig. \ref{DMRGcl}.
Of course, the larger $L$ is the larger $M^*$ is, and it seems that the curves
do not saturate indicating, as is reasonable, that a finite $M$ is not enough
to describe properly a (quasi-)critical system with arbitrarily large $L$.
However, the promising feature of fig. \ref{DMRGcl} is that the various data sets
approximately lie in a strip of the $M-\ln{L}$ plot, characterized by a linear slope
$\bar{M}_0 = 15 \pm 3$. More precisely, this value is an overall measure of the slopes
of the best-fit straight lines, reported in table \ref{m0s} for the six sets considered.
\begin{table}
\caption{\small Slopes $M_0$ of the best fits of the sets plotted in fig. \ref{DMRGcl} (with same notations).
$^\dagger$ Only with $L\geq 28$; $\ddagger$ Only with $L\geq 24$.}
\label{m0s}
\centering{
\begin{tabular}{lc}
\hline\noalign{\smallskip}
Data Set    & $M_0$ \\
\noalign{\smallskip}\hline\noalign{\smallskip}
GS XX & 16.9 $\pm$ 0.4 \\
1st $M^z=0$ & 17.6 $\pm$ 0.4 \\
1st $M^z=1$ $^\dagger$& 15.6 $\pm$ 0.4 \\
\noalign{\smallskip}\hline\noalign{\smallskip}
GS H-D & 13.2 $\pm$ 0.3 \\ 
1st $M^z=0$ & 15.0 $\pm$ 0.2 \\
1st $M^z=1$ $\ddagger$ & 11.4 $\pm$ 0.3 \\
\noalign{\smallskip}\hline
\end{tabular}
}                                 
\end{table}  
In other terms, if we invert the relation between $M$ and $L$, we
find $L^*(M) \propto \exp{(M/M_0)}$, which means that for fixed $M$ we expect a good convergence
of the energies for $L < L^*(M)$ (negligible truncation error, in the language
of \cite{LF}) and that the gain in increasing $M$ is exponential
with a surprisingly small reference value $M_0$ (in line with the first observations of White
himself \cite{Wh93}). Interestingly enough, the logarithmic behavior of $M^*(L)$
can be justified by taking eq. (\ref{wj_vs_epsilon}) to be valid even at critical
points. There one expects $\epsilon \to 0$ and indeed conformal invariance
indicates that the finite-size level spacing vanishes as $\epsilon = \epsilon_0/\ln{L}$
(with $\epsilon_0$ a constant) \cite{PKO}.
Now, if we approximate the discarded weight of eq. (\ref{disc_W}) ignoring the degeneracy factor
$D_j$, we readily get:
\begin{equation}
W_M = \sum_{j > M} Z^j = \frac{Z}{1-Z} Z^M \simeq \frac{\ln{L}}{\epsilon_0}
{\rm e}^{-\epsilon_0 M/\ln{L} } \; .
\label{just_MastL}
\end{equation}

\par
\begin{figure}
\begin{center}
\resizebox{0.45\textwidth}{!}{\includegraphics[keepaspectratio,clip]{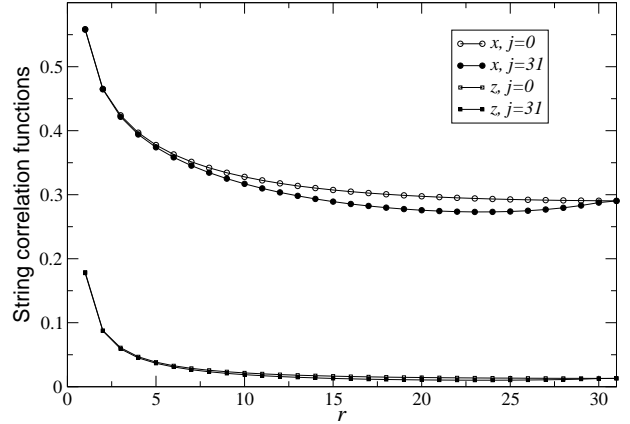}}
\caption{\small Transverse ($\alpha=x$, circles) and longitudinal ($\alpha=z$, squares) 
string correlation functions for a chain
of 64 sites with PBC at the XX point (see text), computed with only $M=50$ DMRG states. 
The empty symbols represent $O_{\rm S}^\alpha(j=0,r)$
while the full ones represent $O_{\rm S}^\alpha(j=31,31+r)$, with an evident dependence on $j$.} 
\label{ntiscf}
\end{center}
\end{figure}

 Unfortunately, the effective accuracy gets poorer, by one or two orders of magnitude
\cite{LRH}, when we deal with correlation functions. Keeping the errors in the low-lying
levels below the desired threshold may not be sufficient so that we had to adopt
an additional criterion to decide whether the selected $M$
is large enough or not. In practice, we check systematically the properties of translational
and reflectional invariance that we expect from the symmetries of the Hamiltonian.
In fact we have observed that of one the ``symptoms'' 
for a too-small $M$ is the visible
(i.e. above numerical uncertainties) lack of some of these invariances.
To be more specific, if $C(0,k)$ is a certain correlation function computed starting at $j=0$,
we have always increased $M$ (at the expense of $L$) until the bound 
$\vert C(L/2,L/2 \pm k)-C(0,k) \vert / \vert C(0,k) \vert \lesssim 0.05$
was met for $k$ varying from $0$ to $L/2$, possibly with the exception of the ranges
where $C(0,k)$ is very small, say $10^{-6}$. 
In fig. \ref{ntiscf} we give an example with string correlation functions
(eq. (\ref{SOPalpha_cf})) at the XX point. These should be translationally invariant
but their numerical estimates depend in fact on the starting point $j$
because we have intentionally fixed a too-small value, $M=50$. It is also interesting to point out
that in this example $O_S^x(j,j+r)$ essentially coincides with $(-)^r \langle S_j^x S_{j+r}^x \rangle$
(not plotted). We interpret this coincidence with the onset of planar order at the BKT transition.
At the XX point both the transverse and the longitudinal string order parameters are expected to vanish 
(sec. II of \cite{AH}), and the fact that the minimum of the $x$-data in fig. \ref{ntiscf} is about $\sim 0.3$
is to be interpreted as a finite-size effect.   
In a similar way, a preliminary study of the 1D Hubbard model with bond-charge interaction 
near its critical points \cite{AM} indicates that in order to obtain accurate correlation functions
it is necessary to use a large number of DMRG states ($> 500$) and several finite-system iterations. From the literature it turns out that this is a general feature of electronic systems, like Hubbard models and critical variants.

\par Finally, we mention the relevance for ref. \cite{GP},
 where the thermal behavior of the quantum 1D and 2D $S=1$ XX model is studied
by means of the two-time Green's function method. The adopted decoupling scheme requires
the calculation of the on-site, nearest-neighbor and next-to-nearest-neighbor ordinary correlations in the
$x$ and $z$ channel:
\begin{displaymath}
C_0^{x,z} \equiv \langle (S_0^{x,z})^2) \rangle \; , \;\; 
C_1^{x,z} \equiv 1/2 [\langle S_0^{x,z} S_1^{x,z} \rangle + \langle S_0^{x,z} S_{L-1}^{x,z} \rangle] 
\end{displaymath}
\begin{equation}
C_2^x \equiv 1/2 [\langle S_0^{x,z} S_2^{x,z} \rangle + \langle S_0^{x,z} S_{L-2}^{x,z} \rangle] \; ,
\label{C_ns}
\end{equation}
(as in eqs. (14) and (15) of \cite{GP} even if it is not clear if $C_2^x$ should include the on-site
term with $\delta=-\delta'$ or not). It is explicitly said that an exact term of comparison in 1D at $T=0$ would
be needed in order to check more precisely their methodology. Though in principle approximate, the
DMRG data can be regarded as a solid benchmark. We have computed the five numbers above with $M=300$
for $L=32,48,64,80,100$. Having to do with short-range non-vanishing quantities, the size dependence
is very weak (fourth decimal place or better) and an algebraic extrapolation to $L \to \infty$
yields $C_0^x=0.7577$, $C_0^z=0.4846$, $C_1^x=0.5579$, $C_1^z=-0.1778$ and $C_2^x=0.464$ (not including
the on-site term). From a direct comparison with table I of \cite{GP} we see that with their
choices of the decoupling parameters the largest
discrepancy affects $C_2^x$ with a relative difference of about 20\%.

\section{Examples and Results}
\label{S_ER}
\par The quality of the numerical analysis of the critical properties depends heavily
on the location of the critical points of interest. 
At present, our study focuses primarily on the transitions from the Haldane phase, for
which it is convenient to fix some representative values of $\lambda$ and let
$D$ vary across the phase boundaries.
This preliminary task of finding $D_{\rm c}(\lambda)$ turns out
to be crucial for subsequent calculations and is divided into two steps.

\par
\begin{figure}
\begin{center}
\resizebox{0.45\textwidth}{!}{\includegraphics[keepaspectratio,clip]{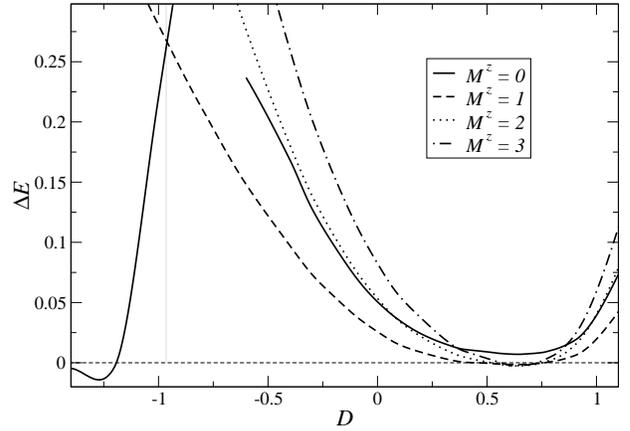}}
\caption{\small Extrapolations of the gaps between the GS and the first excited states
within the indicated sectors of $M^z$ at $\lambda=0.5$. The values for $L \to \infty$ have been
obtained through a quadratic best fit in $1/L$ from the data at $L=32,48,64,80$
with 216 DMRG states. For graphical convenience, the various sets
have been turned into continuous curves using splines. 
The vertical gray line indicates the point of intersection between
the states with $M^z=0$ and $M^z=1$, that is, the analogue of the Haldane triplet at $\lambda=0.5$.}  
\label{Gap_Linf_la05}
\end{center}
\end{figure}

 First, one gets an approximate idea of the transition points using
a direct extrapolation in $1/L$ of the numerical values of the gaps, computed at
increasing $L$ with a moderate number of DMRG states. Clearly, one may want to explore
a rather large interval of values and so the increments in $D$ will not be particularly
small (say 0.1). An example of such a scanning at $\lambda=0.5$ is presented in fig. 
\ref{Gap_Linf_la05}.
Note that at $(\lambda=0.5,D \simeq 0.6)$ 
one has a first insight of the ``cascade'' of levels predicted by CFT
(eq. (\ref{EES}) for $L \to \infty$.)

\par
\begin{figure}
\begin{center}
\resizebox{0.45\textwidth}{!}{\includegraphics[keepaspectratio,clip]{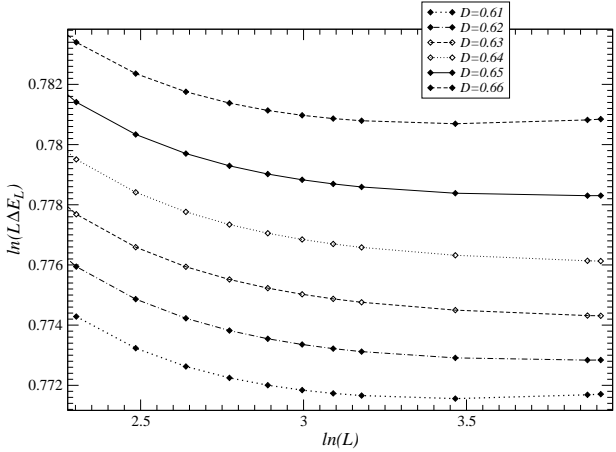}}
\caption{\small H-D transition at $\lambda=0.5$: Scaled gaps 
(between $\vert M^z=0,{\rm b}=0 \rangle$ and $\vert M^z=1,{\rm b}=0 \rangle$)
at $L=10,12,14,16,18,20,22,24,32,48,50$ (with 400 DMRG states), for the values of $D$ indicated in the legend.}  
\label{loglog_la05}
\end{center}
\end{figure}

\begin{figure}
\begin{center}
\resizebox{0.45\textwidth}{!}{\includegraphics[keepaspectratio,clip]{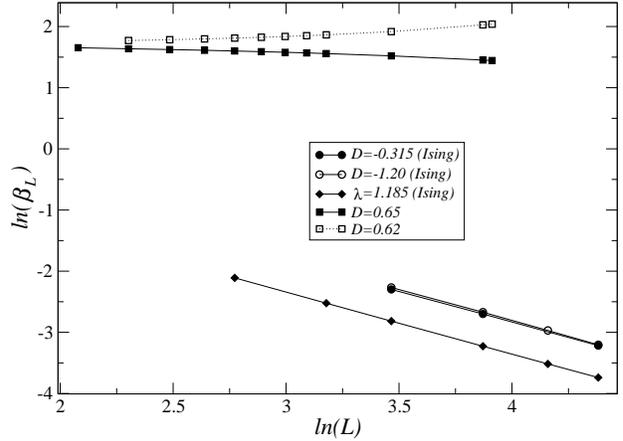}}
\caption{\small Finite-size $\beta$-functions for $(\lambda=0.5,D=0.62)$ (open squares), $(\lambda=0.5,D=0.65)$ (full
squares), $(\lambda=0.5,D=-1.2)$ (open circles), $(\lambda=1,D=-0.315)$ (full circles) and $(\lambda=1.185,D=0)$
(full diamonds).
All the numerical derivatives have been calculated by means of centered differences with $\delta D=0.01$
for the first two cases (Haldane-large-$D$ transition),
$\delta D=0.005$ for the second two and $\delta \lambda=0.005$ for the last one.
Notice that the three lines for the transitions from the Haldane to the N{\'e}el-like phase
(Ising type) have essentially the same slope, that is, the same $\nu$ (see also the discussion
at the end of sec. \ref{S_ER}).}
\label{betaLs}
\end{center}
\end{figure}

 As a second step, the analysis must be refined around the minima of the curves $\Delta E$-vs-$D$
with smaller increments in $D$ and a larger value of $M$. According to FSS 
theory (\cite{HB} and appendix B), at the
true critical point (that is, in the TL) one should see that the data settle to a constant in the log-log plot 
of the scaled gaps vs $L$. Fig. \ref{loglog_la05} shows an example of such an inspection for $\lambda=0.5$
and $D$ varying about the H-D transition. 
It is seen that the differences in the slopes of the various
curves are not so pronounced. Hence, from this plot we have selected two candidates for the
critical point $D_{\rm c}(0.5)$, namely $D=0.62$ and $D=0.65$. At this stage we should mention that
in this type of transition the phenomenological renormalization group (PRG) method 
\cite{HB} also typically yields a pair of (pseudo-)critical values at fixed $L$. In this case, these two sequences of
values seem to converge to the point $D=0.62$ (but in order to see convergence of the curves up to $L=50$
we had to use 400 DMRG states). 
Unfortunately, this turns out to be an invalid tie-break
because at this point the finite-size $\beta$-function (eq. (\ref{nu_from_beta}))
increases with increasing $L$. 
This is shown in fig. \ref{betaLs}, where
$\beta_L(0.65)$ is also plotted. 
The latter scales to zero with a size-dependent
slope that we calculate from eq. (\ref{invnuL}). The extrapolation to $1/L \to 0$ 
(and restricted to $L \geq 22$) then gives
$\nu=3.69 \pm 0.04$. As discussed below this is not a particularly good estimate of $\nu$. Nonetheless, the
location of the critical point $D_{\rm c}(0.5)=0.65$ is quite close to the value $D=0.635$ obtained
in ref. \cite{CHS} with a method based on twisted boundary conditions and exact diagonalization
up to $L=16$. In this sense, we suspect that the difficulties encountered both with FSS and PRG
(roughly speaking, the ``splitting'' of critical points) are due to the peculiar structure
of the energy spectrum in this type of transition. In particular, both sides of the transition are massive
and it is likely that with PBC we are faced with the scenario proposed by Kitazawa \cite{Ki}: In eq.
(\ref{esf}) the constant $C^{(1)}$ for the first excited state could vanish and we have
to consider a second-order expansion in $(D-D_{\rm c})$. Consequently, the values of the critical points
are determined via parabolic intersections that are more sensitive to numerical uncertainties.
Probably this is also the reason why the scaling analysis of $\beta_L$ as used here performs poorly. In fact,
in our framework the best
estimate of $\nu$ comes from another method, namely via the scaling dimension of the mass-generating operator.
According to the discussion that follows, for $(\lambda=0.5,D=0.65)$ we find $\nu=2.38$, essentially
in accord with the values given in \cite{GS}.

\par
\begin{figure}
\begin{center}
\resizebox{0.4\textwidth}{!}{\includegraphics[keepaspectratio,clip]{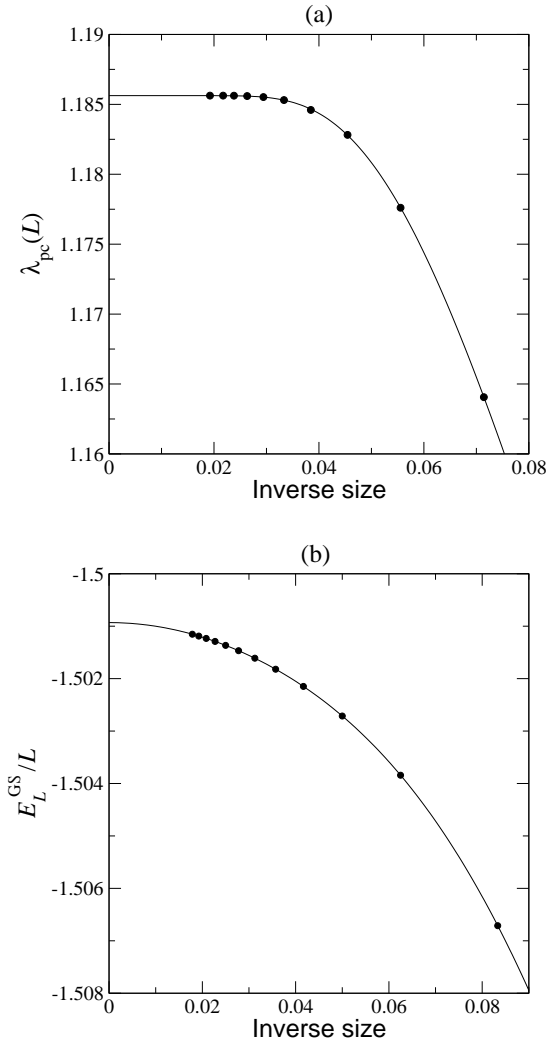}}
\caption{\small (a): PRG location of the Ising-like critical point $\lambda_{\rm c}$ at $D=0$. The pseudo-critical
values of $\lambda$ (full circles) are plotted against the inverse of the middle size, $(L+\delta L/2)^{-1}$,
and correspond to $L$ ranging from 12 to 48 in steps of $\delta L=4$ (see eq. (\ref{PRGeq})) 
using $M=405$. The continuous line
is a best-fit of the form $\lambda_{\rm c}-A L \exp{(-L/L_{\rm PRG})}$ with $A=0.1185$ and $L_{\rm PRG}=3.224$.
(b): Extrapolation (continuous line) of the GS energy density (full circles) according to the form
discussed in the text.}
\label{PRG+EGS_D0Ising}
\end{center}
\end{figure}

 At the transition between the Haldane and the N{\'e}el-like phase 
(henceforth denoted as H-N)
these difficulties regarding the location of the critical points and
the scaling to zero of $\beta_L$ are not experienced. 
Let us work out in detail the $\lambda$-driven transition at $D=0$,
that has been recently revisited \cite{CCC} to argue that it does not belong to the Ising
universality class, as is generally accepted. Taking advantage of previous results,
we fixed $D=0$ in eq. (\ref{HlD}) and varied $\lambda$ about 1.18. In fig. \ref{PRG+EGS_D0Ising}(a)
we report the $L$-dependent pseudo-critical values obtained by solving numerically the PRG equation
\cite{HB}:
\begin{equation}
\frac{[(L+\delta L) \Delta E_{L+\delta L}(\lambda_{\rm pc})-L \Delta E_L(\lambda_{\rm pc})]}{\delta L} = 0 \; .
\label{PRGeq}
\end{equation}
Our best-fit critical value in the TL turns out to be $\lambda_{\rm c}(D=0)=1.1856$ (see caption), 
in agreement both with \cite{KN} and with \cite{CCC}. 
This point separates a gapfull phase, where the scaled gap increases with increasing $L$,
from a gapless one (doubly degenerate GS), where it is expected that the scaled gap converges rapidly
to zero \cite{BJK}. The algebraic decay of $\beta_L$ 
is rather evident from fig. \ref{betaLs} and one
can readily estimate $\nu=0.987 \pm 0.002$ from the slope of the linear best fit.
This is the first indication that the H-N transition belongs to the (2D) Ising universality class,
as reported by various authors (see below our argument based on the combined use of CFT and DMRG). 

\par Despite some difficulties in locating the critical points of the H-D line on the basis of
DMRG data, we are now in position to suggest an extension in the usage of this numerical method
to approach the critical points of quantum Hamiltonians. Taking advantage of the underlying
conformal theory that should be present in all these cases, we focus on the finite-size energy
spectrum, as summarized by eqs. (\ref{EES}) and (\ref{EGS_and_c}).
Once the critical point is located, we select a number of states that seem to become degenerate
with the GS in the limit $L \to \infty$. Then we consider $L \Delta E_L /2 \pi$ and
plot the data against $1/L$ to see whether they settle to a constant, which represents the 
associated scaling dimension
$d_\ell$ multiplied by the velocity $v$. Actually, due to this pre-factor we have to
imagine a self-consistent procedure: Depending on the type of the transition we have in mind
(that is, depending on the conformal anomaly $c$), we stick on one or more levels in the
spectrum that have exactly $d=1$. Hence the value for $1/L \to 0$ is nothing but $v$.
Once the velocity is estimated, one uses eq. (\ref{EGS_and_c}) to best fit the product
$cv$ and see if the value of $c$ and the hypothesis concerning the universality class are self-consistent
or not. 

\par To clarify the matter, let us return to the H-N transition, 
that is thought to be in the universality class of the classical 2D Ising model with $c=1/2$.
This can be considered as the paradigm of minimal models in CFT and the corresponding
quantum field theory is that of a massless Majorana fermion.
Apart from the identity (with
zero scaling dimensions) one has only two primary operators with $\Delta_h=1/16$ and 
$\Delta_T=1/2$. In addition, modular invariance \cite{CaLH} demands that
the conformal spin $(\Delta-\bar{\Delta})$ of the combinations that enter the partition
function (on the torus) must be an integer (zero in this case).
Hence we are left with two nontrivial primary operators of scaling dimensions $d_T=1$
and $d_h=1/8$, that are interpreted as energy and spin density, respectively.
The former is responsible for the variations away from the critical temperature
and so the correlation length index is given by $\nu=1/(2-d_T)=1$.
The latter, from eq. (\ref{O_and_itsd}), gives the decay exponent of the
spin-spin correlation function $\eta_z=1/4$ at the critical point. At the same time,
being $\nu=1$ (see the following eqs. (\ref{ansatzFSSg}) and (\ref{cfOcalQ}) in appendix B), 
one determines also the exponent $\beta=d_h=1/8$, describing the opening of the magnetization 
away from criticality.  

\par
\begin{figure}
\begin{center}
\resizebox{0.45\textwidth}{!}{\includegraphics[keepaspectratio,clip]{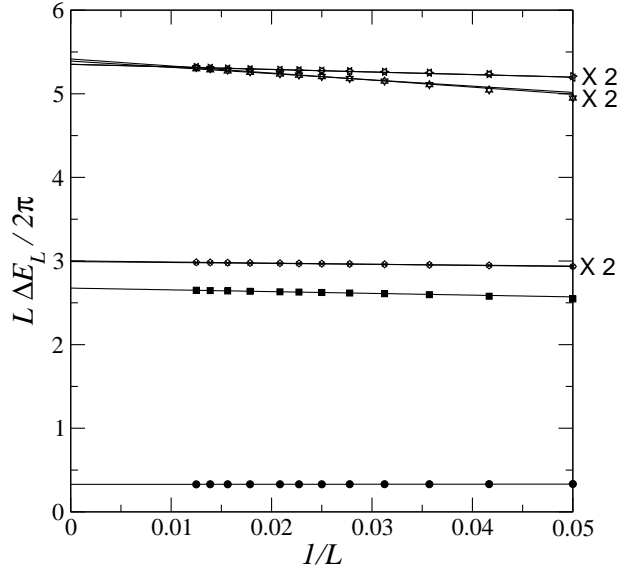}}
\caption{\small Scaled gaps, divided by $2 \pi$, plotted vs $1/L$ at the Ising transition
$(\lambda=1.1856,D=0)$. Points represent the numerical values obtained with
multi-target DMRG runs ($L=20,24,28,32,36,40,44,48,56,64,72,80$ with $M=243$) 
collecting {\it eight} excited states within $M^z=0$.
Continuous lines the are best-fit whose intercepts
are given in table \ref{sdIsing}, together with the theoretical predictions 
of the scaling dimensions (the labels on the right
indicate the multiplicities).}  
\label{SG_D0Ising}
\end{center}
\end{figure}

 The GS energies at various $L$ for the critical point $(\lambda=1.1856,D=0$) found above
are plotted in fig. \ref{PRG+EGS_D0Ising}(b). We have seen that in the range $L=12-56$ the small
deviations from $-1.5$
are accurately fitted by the form $E^{\rm GS}_L/L=e_\infty-cv \pi/6L^2 - A/L^{3/2}
\exp{(-L/\xi_1)}$, with $e_\infty=-1.50093$, $cv=1.327$, $\xi_1=3.48$ and $A=1.251$.
This dependence can be justified by recalling that the lattice model is mapped onto
a CFT (yielding eq. (\ref{EGS_and_c})) for the critical sector, plus a residual Hamiltonian that accounts
for the massive levels. The exponential term is what one expects in a massive regime
\cite{QLY}, and we observe that the length $\xi_1$ multiplied by the finite gap with the sector $\vert M^z=1\vert$,
$\Delta E^{(1)} = 0.754$, yields $v_1 = 2.62$, a typical value of velocity 
(close to the one found below).
In fact, at this stage we observe a nontrivial feature:
the massless modes described by the CFT seem to be associated only with the levels
within $M^z=0$, while those with $M^z \neq 0$ maintain a finite energy gap in the TL.
Hence, the reference state for the calculation of $v$ will be the second excited state in $M^z=0$,
corresponding to conformal dimensions (1/2,1/2).
Using eq. (\ref{v_as_dEdk}) we determine a (CFT) velocity that extrapolates 
to $v=2.676 \pm 0.001$ (see fig. \ref{SG_D0Ising}), and the resulting central charge, $c=0.4959 \pm 0.0006$,
lead us again to the 2D Ising universality class as in \cite{KN} and refs. therein. 
An even more stringent confirmation
comes from fig. \ref{SG_D0Ising}, that illustrates the core of the multi-target method
proposed here.
In table \ref{sdIsing} we compare the theoretical values
with the numerical estimates of the scaling dimensions (second column) obtained 
by taking the ratios between the intercepts at $1/L=0$ of the data set 
in fig. \ref{SG_D0Ising} and the velocity 
pre-factor $v=2.676$. These intercepts, in turn, are determined using power-law fits to eliminate the residual
dependence on $L$ in a fashion similar to that of \cite{CHS_JPSJ}.
The fits are carried out in the range $L \geq 36$ because of the shoulder at $L \sim 40$ in fig. \ref{PRG+EGS_D0Ising}
that signals the onset of the scaling regime.  
All the multiplicities are met, even if with
DMRG calculations alone we are not able to classify the four degenerate states in terms of the secondary
indices. The fourth column contains the
total momentum/conformal spin expected from eq. (\ref{HCFT_Q}).
Question marks indicate that the conformal continuum theory predicts 0 also for those cases
that are expected to have $\vert Q \vert =\pi$ \cite{BJK}. We suspect that this is due to the correspondence between
the original spin model and the field theory that maps the discrete AF structure ($\vert Q \vert=\pi$) 
onto the low-momentum sector. Indeed, for the CFT to give $\vert Q \vert=\pi$ a secondary index equal
to $L/2$ would be needed, and this would yield a nonzero energy gap in the TL.
The overall agreement 
is good (1\% in the worst cases) and the complete structure of the spectrum
of the $c=1/2$ minimal model is reproduced (only the relevant and marginal
cases with $d \leq 2$ are reported). Note that all the marginal operators have nonzero momentum
and so they cannot represent a valid perturbation to the continuum Hamiltonian because
they would break translational invariance. The absence of marginal operators suggests that
each point of the H-N transition corresponds to the same $c=1/2$ theory and the line
in the phase diagram is ``generated'' by the mapping of the discrete spin model onto the
continuum CFT. Moving along the H-N line we expect only a change in the non-universal quantities
like $v$, $e_\infty$ and the values of the critical anisotropies themselves.
We have seen that this is indeed the case, at least for these other two points:
$[\lambda=0.5,D_{\rm c}(0.5)=-1.2]$ (left part of fig. \ref{Gap_Linf_la05}) 
and $[\lambda=1,D_{\rm c}(1)=-0.315]$ (both found by fixing
$\lambda$ and varying $D$). For the former we get $e_\infty=-2.00120$, $v=2.44$ and
$c=0.5008 \pm 0.0008$, 
while for the latter we get $v=2.65$, $e_\infty=-1.62651$ and $c = 0.498 \pm 0.002$. 
The corresponding gap exponents are estimated from
the decay of $\beta_L(-1.2)$ and $\beta_L(-0.315)$, with
the results $\nu=1.023 \pm 0.009$ and $\nu=1.003 \pm 0.006$, respectively. 
Notice that in fig. \ref{betaLs} the two points
have essentially the {\it same} $\beta$-function. After all, eqs. 
(\ref{esf}) and (\ref{nu_from_beta}) give
$\beta_L^{-1}(g_{\rm c})= C^{(1)} L^{1/\nu} \sigma_{\rm c}/(2-\nu^{-1})$, where 
$\sigma_{\rm c}$ is the derivative of $(g-g_{\rm c}$) with respect
to the lattice parameter that is varied. Hence, if in fig. \ref{betaLs} we wish to compare
the $\beta$-functions of Ising type computed at fixed $\lambda$ with those computed
at fixed $D$, we should multiply the latter by the (local) slope of H-N line
${\rm d}D/{\rm d}\lambda \simeq 1.697$. We have checked that with a
vertical shift of $\ln{1.697}=0.5289$ the $\beta$-function at $(\lambda=1.185,D=0)$
collapses onto the other two, thereby indicating that moving along the line
the $c=1/2$ conformal structure is maintained. 
    
\par Concerning the discrepancy with ref. \cite{CCC}, we should mention
that a distinctive point of their analysis is the extrapolation for $M \to \infty$.
On the other hand, we feel that the numerical procedure contains two weak points.
First, the conclusion that $\nu$ is not sufficiently close to 1, and the consequence
that the effective dimensionality is not 2, are drawn from the estimates
of the correlation length using a {\it pure} exponential law 
ignoring the contribution of algebraic prefactors.
According to our experience, having a multi-target code at one's disposal, it would be better
to read the values of $\nu$ directly from the excitation gap or the scaling dimension.
The second and more important point is the usage of the infinite-system algorithm.
In secs. \ref{S_Outline}, \ref{S_MtM} and \ref{S_DepM} we have given indications that
this may be a risky procedure if one aims at very precise quantitative results,
and that the finite-system algorithm should be preferred (see also refs. \cite{LF,TM}
regarding this question). 
\begin{table}
\caption{\small Spectrum of theoretical ($d_{\rm CFT}=\Delta+\bar{\Delta}$+sec. indices)
and numerical (see details in the text) conformal dimensions at the Ising transition of fig. \ref{SG_D0Ising}.}
\label{sdIsing}
\centering{
\begin{tabular}{lccr}
\hline\noalign{\smallskip}
$d_{\rm CFT}$ & $d_{\rm num}$ & Secondary & $Q$       \\
$[\times$ multiplicity] & & Indices & \\
\noalign{\smallskip}\hline\noalign{\smallskip}
0 [$\times 1$]                & & (0,0)         & 0         \\
\noalign{\smallskip}\hline\noalign{\smallskip}
1/8 [$\times 1$]              & $0.1229 \pm 0.0004$ & (0,0)         & 0 (or $\pi$?)\\
\noalign{\smallskip}\hline\noalign{\smallskip}
1 [$\times 1$]                & $1$                 & (0,0)         & 0 (or $\pi$?)\\                         
\noalign{\smallskip}\hline\noalign{\smallskip}
9/8 [$\times 2$]              & $1.1185 \pm 0.0008$   & (1,0)         & $\pm 2\pi/L$  \\
                              & $1.1218 \pm 0.0008$   & (0,1)         & $\mp 2\pi/L$ \\
\noalign{\smallskip}\hline\noalign{\smallskip}
2 [$\times 4$]                & $2.014 \pm 0.005$     & (2,0)         & $\pm 4\pi/L$  \\
                              & $2.025 \pm 0.005$     & (0,2)         & $\mp 4\pi/L$ \\
                              & $2.018 \pm 0.005$     & (1,0)         & $\pm 2\pi/L$  \\
                              & $2.018 \pm 0.005$     & (0,1)         & $\mp 2\pi/L$ \\
\noalign{\smallskip}\hline                              
\end{tabular}
}
\end{table}

\section{Concluding Remarks}
\label{S_CR}

\par The aim of the this paper was to give a detailed explanation of the numerical method
used to derive the results of ref. \cite{DEBEOR} and, more generally, to discuss
how one can circumvent some of the problems of the DMRG close to criticality.
The ideas are explicitly worked out taking some representative point in the $\lambda-D$ phase
diagram for the Hamiltonian of eq. (\ref{HlD}). In sec. \ref{S_DepM} we have argued
that close to the transition lines from the Haldane phase (at least the ones with $c=1$)
the convergence is controlled by a characteristic length $L^*(M)$ that appears to be a consequence
of the truncation to the $M$ states with the largest weights in the block density matrix $\rho_{\rm b}$.
This fact is in line with similar results by other authors \cite{LF,ABO,NOK}, and this probably
lies at the heart of the problem: in the TL the physical system behaves in a critical way but
the DMRG procedure introduces a spurious length so that the numerical outcome is no longer
scale invariant.

\par The method that we are suggesting is based on the finite-system DMRG algorithm
in order to reduce as much as possible the (environment \cite{LF}) errors and has to be
applied in an intermediate range of $L$, not too small so that the signatures of scaling are
visible but not too large as compared to $L^*(M)$. Then we make use of powerful finite-size
scaling predictions, coming from CFT, to extract information on the effective
continuum model, like the ``spin velocity'' and the central charge, from the spectrum
of low-lying excitations. The latter can be computed, within the DMRG, by taking $\rho_{\rm b}$
as the average of the matrices associated with a given number of excited states that, in turn,
are obtained with a Thick-restart variant of the Lanczos method \cite{WS}.

\par We think that the methods discussed here should apply also to the study of the critical
behavior of other 1D quantum systems, such as Hubbard models and their generalizations
\cite{DN,FOPZ,ZJ,LQXCTS}.

\par Finally, in sec. \ref{S_ER} we have reported some examples of $c=1/2$ transitions
between the Haldane and the N{\'e}el-like phase. In particular, we have re-examined
and confirmed the 2D Ising nature of the critical point at $D=0$ \cite{KN}, 
which was the cause of a recent controversy \cite{CCC}.

\section{Acknowledgements}
\label{S_Grazie}
\par We are grateful to L. Campos Venuti, E. Ercolessi, G. Morandi, F. Ravanini, M. Roncaglia and
S.-W. Tsai for their useful interventions during the preparation of the paper. This work
was partially funded by the Italian MIUR, through COFIN projects 
prot. n. 2002024522\_001 and \break 2003029498\_013.

\appendix
\section*{Appendix A: Thick-restart Lanczos Method}
\label{S_TRLM}
\par Let us consider a $n \times n$ real symmetric matrix $H$ representing the Hamiltonian
operator of a quantum system in a given orthonormal basis. We are interested in the lowest
eigenvalues, $\{ \lambda \}$, and corresponding eigenvectors, ${\bf x}$, defined by
the equation $H {\bf x} = \lambda {\bf x}$ ($\lambda$ real, ${\bf x} \in {\mathbb R}^n$).
If the matrix $H$ is large and only a small number of eigenvalues are wanted, a projection-based
method is generally used \cite{Pa,Sa}. In these methods one usually builds orthogonal bases
and then performs the Rayleigh-Ritz projection to extract the approximate solutions. When the matrix is
symmetric, the Lanczos method is the most commonly used algorithm, taking advantage of the fact that
the actual matrix to diagonalize can be cast in tridiagonal form.
\par Another method used for large matrices is the Davidson one (and its variants) for which,
however, the tridiagonal form is not assured. When the Hamiltonian matrix is dominated by
the diagonal elements, as occurs in quantum chemistry, the Davidson-Jacobi gives very good results
\cite{Sa}. This does not necessarily hold for a quantum spin system in the basis of
eigenstates of local $S_j^z$ and the preconditioner (especially during the initial steps
of the infinite-system algorithm) can break some of the symmetries of the Hamiltonian.
Given that the choice of a particular method depends on the physical system under study,
in this appendix we focus on the characteristic points of the Lanczos method used for our
numerical results.

\par Given a starting vector ${\bf r}_0 \in {\mathbb R}^n$, an orthogonal sequence
of vectors ${\bf q}_i \;,\;\; i=1,2,\dots$ is generated recursively from the relations:
\begin{equation}
\left\{
\begin{array}{rcl}
{\bf q}_0 & = & 0 \\
\beta_0   & = & \| {\bf r}_0 \| 
\end{array}
\right.
\label{svLr}
\end{equation}
\begin{equation}
\left\{
\begin{array}{rcl}
{\bf q}_i & = &\frac{{\bf r}_{i-1}}{\beta_{i-1}} \\
\alpha_i & = & \langle {\bf q}_i , H {\bf q}_i \rangle \\
{\bf r}_i & = & H {\bf q}_i - \alpha_i {\bf q}_i - \beta_{i-1} {\bf q}_{i-1} \\
\beta_i & = & \| {\bf r}_i \| \;\; , \;\;\; i=1,2,\dots
\end{array}
\right.
\label{Lr}
\end{equation}
The vectors ${\bf q}_i$, called Lanczos vectors, form an orthonormal basis for the Krylov
subspaces:
\begin{equation}
{\cal K}_m = {\rm span} \left\{ {\bf r}_0, H {\bf r}_0, \dots, H^{m-1} {\bf r}_0 \right\} =
{\rm span} \left\{ {\bf q}_1, \dots, {\bf q}_m \right\} \; ,
\label{Ks}
\end{equation}
and satisfy the relations:
\begin{equation}
H {\bf q}_i = \beta_{i-1} {\bf q}_{i-1} + \alpha_i {\bf q}_i + \beta_i {\bf q}_{i+1} \; ,
\label{Hqi}
\end{equation}
which define a tridiagonal matrix, $T_m$, as a representation of $H$ in ${\cal K}_m$:
\begin{equation}
T_m = \left(
\begin{array}{ccccc}
\alpha_1 & \beta_1 & & & \\
\beta_1 & \alpha_2 & & & \\
& & \ddots & & \\
& & & \alpha_{m-1} & \beta_{m-1} \\
& & & \beta_{m-1} & \alpha_{m} \\
\end{array}
\right)
\label{tridiagH}
\end{equation}
The diagonalization of $T_m$, with $m=1,2,\dots$, gives eigenvalues and eigenvectors,
called Ritz values and Ritz vectors, as approximate solutions to the original problem.
The computation of an eigenvalue with good accuracy often requires a large number of
iterations and on most machines there is not enough memory to store all the Lanczos vectors.
It is necessary to limit the number $m$ of generated vectors and restart the procedure.
Since the algorithm defined by eqs. (\ref{svLr}) and (\ref{Lr}) can start with one
vector ${\bf r}_0$, the usual way is to use the computed Ritz vector, if only one
eigenvalue is wanted. If more than one eigenvalue is desired, we can freeze the converged ones
and combine the other vectors into one starting element \cite{Sa}. 
Other methods (Davidson-Liu or block Lanczos) are based on adding a block of states at each step
but, of course, they require more computational resources. 
Typically, a restarting
scheme needs significantly more iterations to compute a solution, but it saves memory usage.
The algorithm developed by Wu and Simon \cite{WS}, summarized in the following, has the
advantage of adding only one vector at every step, essentially preserving the tridiagonal
structure of the projected matrix.

\par The Thick-restart Lanczos method is based on the following observation: let us consider an
arbitrary element of ${\cal K}_m$:
\begin{equation}
\tilde{\bf{q}} = \sum_{j=1}^m y_j {\bf q}_j \; , \;\; y_j \in {\mathbb R}
\label{eqt}
\end{equation}
from relation (\ref{Hqi}) we have (setting $y_0=0$):
\begin{equation}
H \tilde{\bf{q}} = \sum_{j=1}^m \left( \beta_{j-1} y_{j-1} + \alpha_j y_j + \beta_j y_{j+1} \right)
{\bf q}_j + y_m \beta_m {\bf q}_{m+1} \; ,
\label{Hqt}
\end{equation}
so that the residual, ${\bf r}=y_m \beta_m {\bf q}_{m+1}$, is {\it always parallel to}
${\bf q}_{m+1}$ {\it independently from the vector} $\tilde{\bf{q}}$ in ${\cal K}_m$ {\it and
has the same direction of the residual} ${\bf r}_m$ computed from eq. (\ref{Lr})
and defining ${\bf q}_{m+1}$. This fact suggested to Wu and Simon a suitable restarting scheme.

\par When restarting, one has first to determine an appropriate number of Ritz vectors,
say $k$ (usually greater than the number of wanted eigenvalues):
\begin{equation}
\tilde{\bf{q}}_i = \sum_{j=1}^m q_j y_j^{(i)} \;\; , \; i=1,\dots,k \; ,
\label{eqti}
\end{equation}
corresponding to the Ritz values $\lambda_i$. The chosen Ritz vectors
contain the best approximations available for the wan\-ted eigenvectors.
Since the matrix $T_m$ is symmetric, there is no reason to use
any basis set different from its eigenvectors:
\begin{equation}
T_m {\bf y}^{(i)} = \lambda_i {\bf y}^{(i)} \; , \;\; i=1,\dots,k
\label{Tmyi}
\end{equation}
(We denote with a tilde the quantities after the restart, to distinguish them from 
the corresponding ones before the restart.) Using these $k$ vectors, the projected
matrix of $H$, $\tilde{T}_k$, is diagonal and the vectors satisfy the relations:
\begin{equation}
H \tilde{{\bf q}}_i = \lambda_i \tilde{{\bf q}}_i + y_m^{(i)} \beta_m \tilde{{\bf q}}_{k+1}
\; , \;\; i=1,\dots,k \; ,
\label{Hqti}
\end{equation}
with $\tilde{{\bf q}}_{k+1} = {\bf q}_{m+1}$. So we can enlarge the basis
$\{ \tilde{{\bf q}}_1, \dots , \tilde{{\bf q}}_k \}$ with the vector
$\tilde{{\bf q}}_{k+1}$ and, using the symmetry of the matrix $H$,
the vector $\tilde{{\bf q}}_{k+2}$ can be computed
by the residual:
\begin{equation}
{\bf r}_{k+1} = H \tilde{{\bf q}}_{k+1} - \tilde{\alpha}_{k+1} \tilde{{\bf q}}_{k+1}
-\sum_{j=1}^k \beta_m y_m^{(i)} \tilde{\bf{q}}_j \; .
\label{res_kp1}
\end{equation}
Correspondingly, the matrix $\tilde{T}_k$ is extended by one row and one column
into the matrix $\tilde{T}_{k+1}$. The latter is not tridiagonal as the original
$T_m$ but further steps follow the three-terms relations (\ref{Hqi}), defining
$\tilde{\alpha}_i$ and $\tilde{\beta}_i$ for $i=k+2,k+3,\dots,m$, and at step
$m > k+1$ we have the following form for $\tilde{T}_m$:
\begin{equation}
\tilde{T}_m = \left(
\begin{array}{cccccc}
\lambda_1 & & & \tilde{\beta_1} & & \\
 & \ddots & & \vdots & & \\
& & \lambda_k & \tilde{\beta}_k & & \\
\tilde{\beta}_1 & \dots & \tilde{\beta}_k & \tilde{\alpha}_{k+1} & \tilde{\beta}_{k+1} & \\
& & & \tilde{\beta}_{k+1} & \tilde{\alpha}_{k+2} & \\
& & & & & \ddots \\
\end{array}
\right)
\label{Ttm}
\end{equation}
with $\tilde{\beta}_i = \beta_m y_m^{(i)} \; , \;\; i=1,\dots,k$.
Since the vectors in the new subspace generated by $\{ \tilde{{\bf q}}_1, \dots, \tilde{{\bf q}}_m \}$
have the same property that the residual is always parallel to $\tilde{{\bf q}}_{m+1}$,
the procedure can be repeatedly restarted, until we find a good convergence of the
wanted eigenvalues. The matrix $T_m$ is no longer tridiagonal after the first restart, but it can
still be stored and diagonalized in an efficient way. It is easy to arrange the algorithm so that
${\bf q}_i$ and $\tilde{{\bf q}}_i$ occupy the same memory locations.

\par We can summarize saying that the restarting scheme with $k$ Ritz vectors ${\bf q}_i$
(we now drop the tilded notation), and a residual ${\bf r}_k$
satisfying the rule:
\begin{equation}
H {\bf q}_i=\lambda_i {\bf q}_i+\beta_i {\bf q}_{k+1} \; ,
\label{Hqiwlambda}
\end{equation}
with ${\bf q}_{k+1}={\bf r}_k/\| {\bf r}_k \|$, is composed by an initialization:
\begin{equation}
\left\{
\begin{array}{rcl}
\alpha_{k+1} & = & \langle {\bf q}_{k+1} , H {\bf q}_{k+1} \rangle \\
{\bf r}_{k+1} & = & H {\bf q}_{k+1}-\alpha_{k+1} {\bf q}_{k+1} - \sum_{j=1}^k \beta_j {\bf q}_j \\
\beta_{k+1} & = & \| {\bf r}_{k+1} \|
\end{array}
\right.
\label{rLr}
\end{equation}
followed by typical Lanczos iterations (eq. (\ref{Lr})) for $i=k+2,k+3,\dots,m$.

\par This concludes our description of the Thick-restart Lanczos algorithm.
We refer to the original paper \cite{WS} for a detailed discussion of the errors
and precision of the method.

\section*{Appendix B: Brief Survey of FSS Theory}
\label{S_FSS}
\par Ideally, we should locate the critical points of a quantum system by looking
at the smallest energy gap, $\Delta E$, as a function of a (relevant) parameter
$g$ and identify $g_{\rm c}$ through the condition $\Delta E(g_{\rm c})=0$.
For an algebraic transition the critical index $\nu$ controls the opening of the gap, 
$\Delta E \propto \vert g-g_{\rm c}\vert^\nu$.  
However, when the critical point is approached numerically we encounter two related problems.
First, the system's size is necessarily finite and a true phase transition is forbidden. Second,
if we are able to deal with sufficiently large systems close to $g_c$, the energy gap may become
so small as to be comparable with the errors introduced by the algorithm (or, ultimately, by the machine).
Hence we need a prescription to infer the location of the critical point from nonzero values
of $\Delta E_L$ at finite $L$.

\par In FSS theory \cite{GS,HB,FB} one usually invokes the following ansatz
(quantum Hamiltonian notations):
\begin{equation}
\Delta E_L = \frac{1}{L} F(\zeta) \; , \;\; \zeta \equiv L^{1/\nu} \vert g-g_{\rm c} \vert \; ,
\label{ansatzFSS}
\end{equation}
so that the scaling variable $\zeta$ covers these two regimes
\begin{itemize}
\item[(a)] $\zeta \to 0$ for $g \to g_{\rm c}$ 
at fixed $L$. In this regime we should see a finite system at the infinite-size critical
point. Since $\Delta E_L$ is in fact an inverse correlation length $\xi_L$, we expect the latter
to be as large as possible, that is to say $\Delta E_L \propto L^{-1}$. Compatibility
with eq. (\ref{ansatzFSS}) then requires $F(0) \neq 0$. This is exactly the case for which
the continuum description of CFT applies (sec. \ref{S_IdCFT}).
Hence, recalling that the scaling dimension of the gap-generating operator is $(2-\nu^{-1})$, we may 
write $F(0)=2 \pi v (2-\nu^{-1})$ and expand $F(\zeta)$ in a McLaurin series for $\zeta \ll 1$:
\begin{equation}
\Delta E_L = \frac{2 \pi v}{L} [(2-\nu^{-1}) + C^{(1)} \zeta + C^{(2)} \zeta^2 + \dots] \; .
\label{esf}
\end{equation}
\item[(b)] $\zeta \to \infty$, which means $(L/\xi_L)^{1/\nu} \gg 1$. This regime mimics
the TL, in the sense that $\xi_L$ is large but finite because the system is slightly off-critical
and $L$ is sufficiently large so that scaling laws appear. Hence $L$ must effectively cancel
in eq. (\ref{ansatzFSS}), leaving just $\vert g-g_{\rm c}\vert^\nu$. This is possible if:
\begin{equation}
F(\zeta) \sim \zeta^\nu \;\; \zeta \gg 1 \; .
\label{ocr}
\end{equation}
\end{itemize}

\par From (a) we derive the way to locate the critical points through FSS: Plot
$\ln{(\Delta E_L})$ vs $\ln{L}$ and look for the value of $g$ that best gives
a straight line with slope $-1$. Actually, we can be even more severe by looking for slope 0
in the curves of the scaled gaps $L \Delta E_L$.
As far as the index $\nu$ is concerned, one may consider
the finite-size $\beta$-function,
$\beta_L^{-1}(g) \equiv \partial \ln{(\Delta E_L)}/\partial g$, evaluated at $g=g_{\rm c}$
as determined above (apart from the sign):
\begin{equation}
\beta_L(g_{\rm c}) = [{F(\zeta)}/{F'(\zeta)}]_{\zeta=0} L^{-1/\nu} \; ,
\label{nu_from_beta}
\end{equation}
where $F'$ denotes the derivative with respect to $\zeta$ and $F'(0)$ is also assumed to
be nonzero.
In principle $\beta_L(g_{\rm c})$ should vanish with an exponent $1/\nu$
that represents the asymptotic slope in log-log scales. Alternatively,
since the scaling region may be reachable only for very large $L$, one can calculate
the discrete logarithmic derivative through a size increment $L \to L+\delta L$:
\begin{equation}
\frac{1}{\nu_L} \equiv - \frac{\ln{\beta_{L+\delta L}(g_{\rm c})}-\ln{\beta_L(g_{\rm c})}}
{\ln{(L+\delta L)}-\ln{L}} \; ,
\label{invnuL}
\end{equation}
and this should converge to $1/\nu$ when $L \to \infty$.

\par Finally, the ansatz (\ref{ansatzFSS}) can be generalized to other physical quantities
that behave as ${\cal Q}(g) \propto \vert g - g_{\rm c} \vert^{\nu_{\cal Q}}$ near the critical point:
\begin{equation}
{\cal Q}_L(g) = L^{-z_{\cal Q}} F_{\cal Q}(\zeta) \; ,
\label{ansatzFSSg}
\end{equation}
with the same scaling variable $\zeta$ and scaling exponent $z_{\cal Q}$. As above, in the
critical regime (a) $\zeta \to 0$ we shall require $F_{\cal Q}(0) \neq 0$, while
in the off-critical regime (b) the scaling exponent $\nu_{\cal Q}$ emerges provided that
$F_{\cal Q} \sim \zeta^{\nu_{\cal Q}}$ for $\zeta \to \infty$, and $z_{\cal Q}=\nu_{\cal Q}/\nu$. 
In many cases, ${\cal Q}^2$ is a squared order parameter given by the asymptotic value
of a certain correlation function, $\langle O_{\cal Q}(0) O_{\cal Q}(r) \rangle$, that slightly
away from the critical point behaves as:
\begin{equation}
\langle O_{\cal Q}(0) O_{\cal Q}(r) \rangle \propto \frac{G_{\cal Q}(r/\xi)}{r^{2 d_{\cal Q}}} \; ,
\label{cfOcalQ}
\end{equation}
$d_{\rm Q}$ being the scaling dimension of $O_{\cal Q}$. At this stage we can make the scaling
argument of Ginsparg (sec. 5.1 of \cite{Gi}) evaluating eq. (\ref{cfOcalQ}) just at the correlation
length itself, $r=\xi \propto \vert g-g_{\rm c} \vert^{-\nu}$ thereby having
$\langle O_{\cal Q}(0) O_{\cal Q}(\xi) \rangle \sim {\cal Q}^2 \sim 
G_{\cal Q}(1) \vert g-g_{\rm c}\vert^{2 \nu d_{\cal Q}}$.
Hence we find the scaling law $d_{\cal Q}=\nu_{\cal Q}/\nu$, that tells us that $z_{\cal Q}$ in eq. (\ref{ansatzFSSg})
is nothing but the scaling dimension of the operator associated with the order parameter ${\cal Q}$.
In ref. \cite{DEBEOR} we have used this property to derive the decay exponents of
ordinary (transverse channel) and string ($z$-channel) 
correlation functions in $c=1$ phases by applying eq. (\ref{ansatzFSSg})
to the corresponding N{\'e}el and string order parameters evaluated at half chain.

\end{document}